\documentclass[twocolumn,showpacs,superscriptaddress]{revtex4-1}
\usepackage{graphicx}
\usepackage{units}
\usepackage{multirow}
\usepackage{bigstrut}
\usepackage{overpic}
\usepackage{array}
\usepackage{color}
\usepackage{amsmath}
\RequirePackage{lineno}

\begin{document}%%

%\modulolinenumbers[2]
%\linenumbers

\setlength{\oddsidemargin}{-0.5cm} \addtolength{\topmargin}{15mm}

\title{\boldmath Experimental realization of secure multi-party computation \\
in an entanglement access network}

\author{
\small
\begin{center}
X.-Y. Chang$^{1}$, D.-L. Deng$^{2,1}$, X.-X. Yuan$^{1}$, P.-Y. Hou$%
^{1}$, Y.-Y. Huang$^{1}$, L.-M. Duan$^{2,1}$
\\
\vspace{0.2cm} {\it
$^{1}$Center for Quantum Information, IIIS, Tsinghua University, Beijing 100084, PR China \\
$^{2}$Department of Physics, University of Michigan, Ann Arbor,
Michigan 48109, USA \\}\end{center} \vspace{0.4cm} }

\begin{abstract}
To construct a quantum network with many end users, it is critical
to have a cost-efficient way to distribute entanglement over different
network ends. We demonstrate an entanglement access network, where the
expensive resource, the entangled photon source at the telecom wavelength
and the core communication channel, is shared by many end users. Using this
cost-efficient entanglement access network, we report experimental
demonstration of a secure multiparty computation protocol, the
privacy-preserving secure sum problem, based on the network quantum
cryptography.

\end{abstract}

\maketitle

The peculiar quantum correlation, entanglement, provides the crucial
resource for quantum information processing \cite{1,2,3}. Generation of
remote entanglement is a key step for quantum communication and distributed
computation \cite{4,5}. Distribution of entanglement in a multiple-end
quantum network is costly through the point-to-point protocol as one needs
to establish an entanglement source and a quantum communication channel
between each end \cite{4,5,6,7,8,9,10}. We demonstrate a cost-efficient
entanglement access network where multiple end users share the same
entanglement source and the core communication channel. We verify
entanglement and quantum nonlocality between different end users. Using this
entanglement access network, we report the experimental demonstration of a
secure multi-party computation protocol where several end users compute
cooperatively to solve a joint problem without revealing the information of
their actual inputs \cite{11,12,12a}. This demonstration opens up the
prospect of applying the cost-efficient entanglement access network for
achieving secure multi-party computation that protects the privacy of end
users.

With popularity of the internet, distributed computation becomes
increasingly more important, where a number of end users need to work
cooperatively to solve a common problem. The answer to the problem typically
depends on the inputs of all the end users, which need to be communicated in
the network. At the same time, the users need to protect their privacy and
do not want to reveal their information to the others. Secure multi-party
computation is a branch of cryptography and computer science that studies
this kind of problems \cite{11,12,12a}. A well-known primitive example of
this field is the millionaire problem, first introduced by Yao \textbf{\cite%
{11}}, in which two millionaires want to know which of them is richer
without revealing their actual wealth. Secure multi-party computation has
many applications in e-commerce and data mining where people need to compare
numbers which are confidential \cite{11,12,12a}. Classically, the secure
multi-party computation protocols typically rely on computationally hard
mathematical problems, which require specific assumptions and are subject to
security loopholes in particular under attack by a quantum computer \cite%
{12a}.

In this paper, we demonstrate an entanglement access network which offers an
alternative route for secure multi-party computation based on the network
quantum cryptography. Quantum access network is a concept introduced in a
recent paper where the expensive quantum resource is shared by many end
users \textbf{\cite{8,9}}. For quantum key distribution based on the BB84
protocol \textbf{\cite{13}}, the photon detector is the relatively expensive
part of its implementation and thus shared in the quantum access network
demonstrated in Ref. \textbf{\cite{8}}. Here, we realize an entanglement
access network to efficiently distribute entanglement between network end
users. In this network, the entanglement source is the most expensive part
of the implementation and thus shared between many end users. Entanglement
provides the crucial quantum resource for achieving device-independent
quantum cryptography, the most secure way for cryptographic communication
\textbf{\cite{15,16,18,20,21,23}}. Entanglement is also the critical
resource for certified generation of shared randomness \textbf{\cite{23a}},
and for quantum communication and multi-party computation through the
entanglement-based schemes \textbf{\cite{3,24}}. We demonstrate an
entanglement access network which efficiently distributes entanglement
between a number of end users connected through fibers of more than $20$ km
length by sharing a single entangled photon source at the telecom frequency.
By use of this entanglement access network, we report experimental
demonstration of a secure multi-party computation protocol, the secure sum
problem, in which several millionaires want to know how much money they have
in total but none of them is willing to reveal his wealth to others \textbf{%
\cite{11,12}}. This demonstration shows that the entanglement access network
provides a cost-efficient way to realize a quantum network with shared
resource, opening up the prospect for its applications in secure multi-party
cryptography and distributed quantum computation.

The entangled photon source shown in Fig. 1(a) is generated by a type-II BBO
crystal pumped by ultrafast laser pulses (with the pulse duration less than $%
150$ fs and a repetition rate of $76$ MHz) at the wavelength of $775$ nm
from a Ti:Sapphire laser. The spontaneous parametric down conversion in the
BBO crystal produces entangled photon pairs at the telecom wavelength of $%
1550$ nm, which, in the ideal case, are in the polarization entangled state $%
\left\vert \Psi \right\rangle =\left( \left\vert HV\right\rangle
+e^{i\varphi }\left\vert VH\right\rangle \right) /\sqrt{2}$, where $%
\left\vert H\right\rangle $ and $\left\vert V\right\rangle $ denote
respectively the horizontal and the vertical polarization state and $\varphi
$ is a controllable phase \textbf{\cite{24a}}. We have verified that the
experimentally generated state $\rho $ has an entanglement fidelity $F_{s}$
of $\left( 95.12\pm 0.25\right) \%$ with respect to this ideal state $%
\left\vert \Psi \right\rangle $ through the quantum state tomography \textbf{%
\cite{24a2}}.

The entangled photons at the telecom frequency are coupled into optical
fibers representing the core communication channel with the total distance
about $20$ km. The output photons from the fibers are fed into a $1\times 8$
optical switch at each side (called Alice's and Bob's side) which is
electrically controlled by the computer to deliver the photons to one of the
eight output ports according to the electric control signal. This forms an
entanglement access network with $8\times 8$ possible choices of pairs of
end users who can share entangled photons with the entanglement distance
more than $20$ km, as shown in Fig. 1(b). All the end users share the same
entanglement source and the core communication channel. The polarization
states of photons are subject to tension and temperature dependent rotations
in the optical fibers which are carefully compensated afterwards through the
fiber polarization controllers.

\begin{figure}[tbp]
\includegraphics[width=8.5cm,height=10cm]{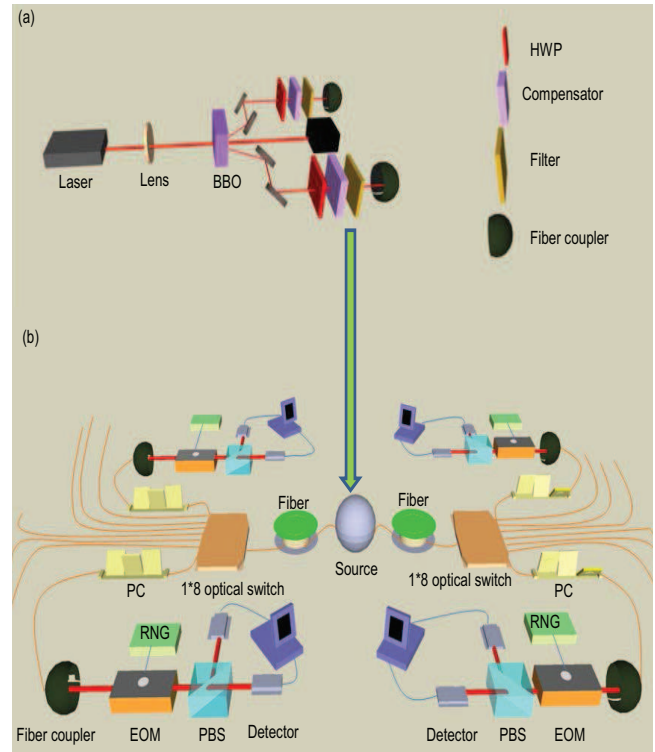}
\caption[Fig. 1 ]{ \textbf{\ The experimental setup for implementation of an
entanglement access network.} (a) The setup to generate the polarized
entangled photons at the telecom wavelength. The ultrafast pulse at the $775$
nm wavelength from the Ti:Sapphire laser is $H$-polarized and focused by
lens onto the BBO crystal cut in the Type-II phase-matching condition,
generating polarized entangled photons at the $1550$ nm wavelength. To
ensure entanglement, we use a compensator made by another BBO crystal at
each output, which compensates the temporal walk-off between the $H$ and $V$
polarized photons in the nonlinear crystal. The output photons are filtered
by $3$ nm interference filters and then coupled into long-distance optical
fibers with a fiber coupler. The half-wave plates (HWP) are used for
alignment of the polarization axes before and after the fibers. (b) The
setup to construct an entanglement access network , where the entanglement
source and the core communication channel are shared by many end users.
Computer controlled optical switches are used to distribute the entangled
photons to different end users. Our experiment demonstrates an entanglement
access network with up to $8\times 8$ end users, where the two sides are
separated by a fiber about $20$ km. Among the end users, we have verified
entanglement and quantum nonlocality between four pairs of them and used the
shared entanglement to demonstrate the four-party secure sum protocol as an
example to illustrate its application for implementation of secure
multi-party computation.}
\end{figure}

To demonstrate entanglement between different end users, we randomly choose
two users $A_{1}$ and $A_{2}$ at Alice's side and two users $B_{1}$ and $%
B_{2}$ at Bob's side. For the four pairs of end users $A_{i}B_{j}$ $\left(
i,j=1,2\right) $, we measure their entanglement by detecting the photon
coincidences in different polarization bases. To rotate the polarization
basis, we use a fast switchable electric optical modular (EOM), which
induces a polarization transformation $\left\vert H\right\rangle \rightarrow
\cos \theta \left\vert H\right\rangle -i\sin \theta \left\vert
V\right\rangle $ and $\left\vert V\right\rangle \rightarrow -i\sin \theta
\left\vert H\right\rangle +\cos \theta \left\vert V\right\rangle $ with the
angle $\theta $ determined by the computer controlled electric signal. The
output $H $ and $V$\ polarized photons are split by a polarization beam
splitter (PBS) and then detected through single photon counters. In the Z
(or X) basis detection, we register the photon coincidence counts by fixing
the angle $\theta _{A}$ of $A_{i}$ at $0^{o}$ (or $45^{o}$) and rotating the
angle $\theta _{B}$ of $B_{j}$. The resulting coincidence counts as
functions of the angle $\theta _{B}$ are shown in Fig. 2 for different pairs
$A_{i}B_{j}$. The big contrast of these oscillations is a clear
demonstration of quantum entanglement. Quantitatively, we calculate the
visibilities of these oscillation curves $V_{z}$ and $V_{x}$ in the Z and
the X basis, respectively, and the entanglement fidelity $F_{e}$ is then
bounded by $F_{e}\geq (V_{z}+V_{x})/2$ \textbf{\cite{25}}. The visibilities
and the corresponding fidelity bounds are listed in Fig. 3. All the pairs
have the entanglement fidelity higher than $90\%$. The small decrease of the
entanglement fidelity compared with the fidelity $F_{s}$ of the entangled
photon source is due to the imperfection in compensation of the polarization
rotation in the optical fibers.

\begin{figure}[tbp]
\includegraphics[width=8.5cm,height=6cm]{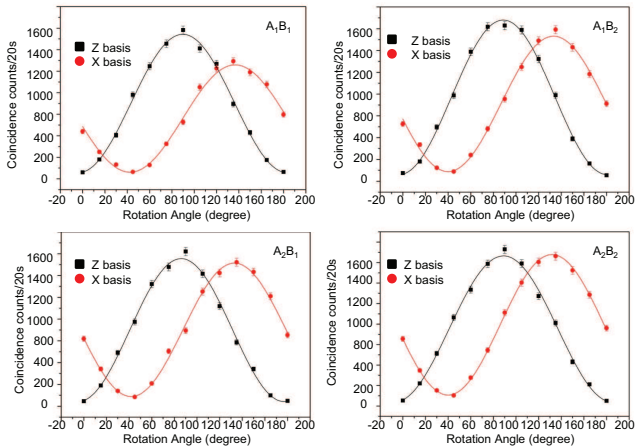}
\caption[Fig. 2 ]{ \textbf{\ The entanglement demonstrated between different
pairs of the end users.} The figures show the coincidence counts when we fix
the rotation angle of the electric optical modulator (EOM) of one end at $%
0^{o}$ (for the Z-basis) or $45^{o}$ (for the X-basis) while rotating the
angle of the EOM at the other end. Different sub-figures correspond to
different pairs of parties ($A_1$ and $B_1$, $A_1$ and $B_2$, $A_2$ and $B_1$%
, $A_2$ and $B_2$). The visibilities of the oscillations in the X and the Z
bases together bound the entanglement fidelity of the photonic states shared
between the corresponding parties. }
\end{figure}

The shared entanglement allows demonstration of quantum key distribution
between any pair of the end users in this network using the Ekert protocol
\textbf{\cite{2}}. For this purpose, we need to randomly choose the angle $%
\theta _{A}$ from the set $\left\{ 0^{o},22.5^{o},45^{o}\right\} $ and $%
\theta _{B}$ from the set $\left\{ 22.5^{o},45^{o},67.5^{o}\right\} $, and
record the individual measurement outcomes for each coincidence count
\textbf{\cite{26a}}. From the counts for the angles $\theta _{A}=\left\{
0^{o},45^{o}\right\} $ and $\theta _{B}=\left\{ 22.5^{o},67.5^{o}\right\} $,
we calculate the expectation value of the Clauser-Horne-Shimony-Holt (CHSH)
observable $\left\langle S\right\rangle =\left\langle
0^{o},22.5^{o}\right\rangle +\left\langle 45^{o},22.5^{o}\right\rangle
+\left\langle 45^{o},67.5^{o}\right\rangle -\left\langle
0^{o},67.5^{o}\right\rangle $ \textbf{\cite{26}}, where $\left\langle \theta
_{A},\theta _{B}\right\rangle $ denotes the photon coincidence counts with $%
\theta _{A},\theta _{B}$ at the specified values. The CHSH values are listed
in Fig. 3 for different pairs $A_{i},B_{j}$. The CHSH inequality $%
\left\langle S\right\rangle \leq 2$ for any classical correlation is clearly
violated, demonstrating quantum nonlocality. The significant violation of the
CHSH inequality for each pair of the end users is a guarantee of the security of the
entangled-based QKD protocol \textbf{\cite{15,16,18,20,21,23}}. To generate quantum key, the
measurement outcomes at $\theta _{A}=\theta _{B}=$ $\left\{
22.5^{o},45^{o}\right\} $ are kept, which yield the sifted keys \textbf{\cite%
{26a}}. For each pair of the parties, we obtain roughly $1.2\times 10^{4}$
sifted keys. We randomly choose $20\%$ of the keys to estimate the quantum
bit error rate (QBER). For our data, the QBER $\gamma $ is about $5\%$,
smaller than the security threshold of $11\%$ \cite{23}. We use the low
density parity check (LDPC) code \cite{27,28} to do the error correction for
the sifted keys, which is more efficient compared with the conventional
CASCADE protocol \textbf{\cite{26a}}. We obtain about $8000$ error-free
shared keys, which are then purified by the privacy amplification protocol
via a universal-$2$-class hash function \cite{29}, yielding about $1800$
shared secret final keys. The residual information available to any
potential eavesdroppers is reduced by a factor of $2^{-300}$ during the
privacy amplification and thus much less than one bit.

\begin{figure}[tbp]
\includegraphics[width=8.5cm,height=4cm]{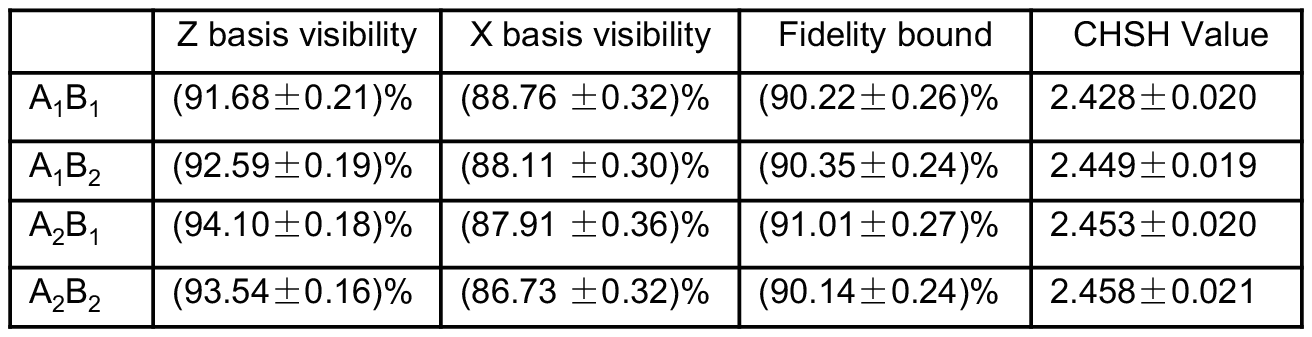}
\caption[Fig. 3 ]{ \textbf{\ The experimental data to show entanglement and
quantum nonlocality shared between different parties.} The table lists the
oscillation visibilities of the coincidence counts in the Z and the X bases,
whose average gives a lower bound to the entanglement fidelity (the fourth
column). The error bars are obtained by assuming a Poissionian distribution
for the photon counts and propagated from the measured coincidence counts to
the quantities listed in the table through exact Monte Carlo simulation. The
last column of the table shows the measured value for the CHSH observable,
and a larger-than-$2$ value indicates violation of the Bell inequality,
demonstrating quantum nonlocality shared between the corresponding parties. }
\end{figure}

We now use the entanglement access network to demonstrate a secure
multi-party computation protocol: the privacy-preserving secure sum problem
\textbf{\cite{12}}. The problem can be illustrated with the following
example: assume several millionaires want to know how much money they have
in total, but none of them want to reveal his actual wealth to others. To be
concrete, we consider the case of four parties $A_{1},A_{2},B_{1},B_{2}$.
Denote by $a_{1}$, $a_{2}$, $b_{1}$ and $b_{2}$ the input from $A_{1}$, $%
A_{2}$, $B_{1}$, $B_{2}$, respectively. We want to calculate the sum $%
T=a_{1}+a_{2}+b_{1}+b_{2}$ without revealing the inputs $%
a_{1},a_{2},b_{1},b_{2} $. The quantum protocol to accomplish this task
using the entanglement access network goes as follows:

\textit{Step 1.}---Using the entanglement-based network QKD, $A_{1}$ and $%
B_{1}$, $B_{1}$ and $A_{2}$, $A_{2}$ and $B_{2}$, and $B_{2}$ and $A_{1}$
share random keys of $n$ bits denoted by $\mathcal{R}_{a_{1}b_{1}}$, $%
\mathcal{R}_{a_{2}b_{1}}$, $\mathcal{R}_{a_{2}b_{2}}$, and $\mathcal{R}%
_{a_{1}b_{2}}$, respectively. The number of bits $n$ is taken to be at least
as large as the estimated number of bits for the sum $T$.

\textit{Step 2}.---$A_{1}$ calculates $X_{1}=a_{1}+\mathcal{R}_{a_{1}b_{1}}-%
\mathcal{R}_{a_{1}b_{2}}$ and publicly announces $X_{1}$ to others; $B_{1}$
calculates $X_{2}=b_{1}+\mathcal{R}_{a_{2}b_{1}}-\mathcal{R}_{a_{1}b_{1}}$
and publicly announces $X_{2}$ to others; $A_{2}$ calculates $X_{3}=a_{2}+%
\mathcal{R}_{a_{2}b_{2}}-\mathcal{R}_{a_{2}b_{1}}$ and publicly announces $%
X_{3}$ to others; $B_{2}$ calculates $X_{4}=b_{2}+\mathcal{R}_{a_{1}b_{2}}-%
\mathcal{R}_{a_{2}b_{2}}$ and publicly announces $X_{4}$ to others.

\textit{Step 3}.---All of them know the sum $T$ by calculating $%
T=X_{1}+X_{2}+X_{3}+X_{4}$. At the same time, the inputs $%
a_{1},a_{2},b_{1},b_{2}$ remain confidential to the other parties as the
public information $X_{i}$ ($i=1,2,3,4$) has no correlation with the inputs
due to the one-time pad theorem. For security of this secure sum protocol,
we have assumed that different parties do not collaborate to steal the input
information of the other parties. With the sum $T$, when a few parties
collaborate, it is always possible to reveal the input information of the
other party.

In the experimental demonstration, as an example, we take randomly generated
numbers $a_{1}=55406$, $b_{1}=116559$, $a_{2}=988150$ and $b_{2}=2839885$.
We run the above implementation of the secure sum protocol $30$ times, and
for each time we use different keys $\mathcal{R}_{a_{1}b_{1}},\mathcal{R}%
_{a_{2}b_{1}},\mathcal{R}_{a_{2}b_{2}},\mathcal{R}_{a_{1}b_{2}}$ with the
number of bits $n=25$ bits to encode the public information $%
X_{1},X_{2},X_{3},X_{4}$. For these $30$ experimental runs, the publicly
announced numbers $X_{1},X_{2},X_{3},X_{4}$ are shown in Fig. 4, which look
completely random from trial to trial and reveal no information of the
inputs $a_{1},a_{2},b_{1},b_{2}$. However, for each run, their sum is always
fixed to be $4\times 10^{6}$, which gives the correct calculation result for
$T=a_{1}+a_{2}+b_{1}+b_{2}$.

\begin{figure}[tbp]
\includegraphics[width=8.5cm,height=10cm]{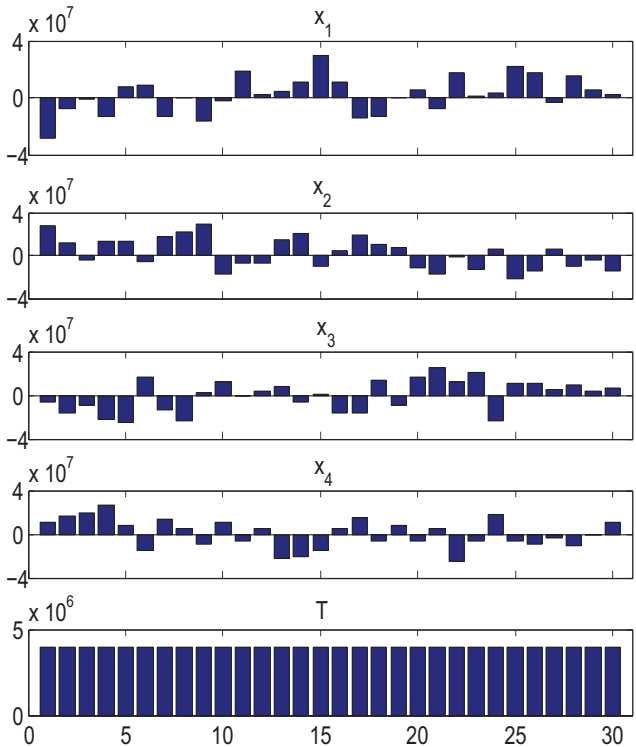}
\caption[Fig. 3 ]{ \textbf{\ The experimental demonstration of the
four-party secure-sum protocol through the entanglement access network.} The
top four sub-figures show the publically announced data from each of the
four parties for $30$ rounds of experiments for demonstration of the same
secure-sum problem with the same input from each party. The randomly
distributed data over different experimental runs is an indication that the
announced data reveal no information of the input from each party. The last
sub-figure shows the secure sum calculated by each party from the publically
announced data, which is identical for different experimental runs and
always equal to the true value of the sum for the underlying problem. }
\end{figure}

Similar to the quantum access network realized recently \textbf{\cite{8}},
we expect that the entanglement access network demonstrated in this paper
provides a source-efficient way for network cryptography and secure
multi-party computation. In particular, the shared entanglement between each
ends of the network opens up the possibility to realize device independent
quantum cryptography \textbf{\cite{15,16,18,20,21}}, which allows most
secure communication by closing the practical security loopholes in
conventional quantum cryptography \textbf{\cite{23}}. Apart from the example
demonstrated in this paper, the entanglement access network may allow
realization of a number of secure multi-party computation problems \textbf{%
\cite{12a}}. The entanglement shared in the network could also find
applications for certified generation of random numbers \textbf{\cite{23a}}
and for implementation of distributed quantum computation and multiparty
quantum cryptography protocols \textbf{\cite{1,3}}.

This work was supported by the National Basic
Research Program of China 2011CBA00302. LMD and DLD acknowledge in addition
support from the IARPA MUSIQC program, the AFOSR and the ARO MURI program.

\end{document}